\documentclass{rspublic}
\usepackage[dvips]{graphicx}
\usepackage{epsfig,epsf}
\usepackage{natbib}

\begin{document}

\title{An optimal path to transition in a duct}
\author[D. Biau and A. Bottaro]{Damien Biau and Alessandro Bottaro}
\affiliation{DICAT, University of Genova, 1 via Montallegro, Genova, Italy}
\label{firstpage}

\maketitle
\begin{abstract}{duct flow; optimal perturbations; minimal defects; transition to turbulence}
This paper is concerned with the transition of the laminar flow in a duct of square cross-section. Like in the similar case of the pipe flow, the motion is linearly stable for all Reynolds numbers, rendering this flow a suitable candidate for a study of the 'bypass' path to turbulence. It has already been shown \citep{Biau_JFM_2008} that the classical linear optimal perturbation problem, yielding optimal disturbances in the form of longitudinal vortices, fails to provide an 'optimal' path to turbulence, i.e. optimal perturbations do not elicit a significant nonlinear response from the flow.  Previous simulations have also indicated that a pair of travelling waves generates immediately, by nonlinear quadratic interactions, an unstable mean flow distortion, responsible for rapid breakdown.  By the use of functions quantifying the sensitivity of the motion to deviations in the base flow,  the 'optimal' travelling wave associated to its specific defect is found by a variational approach. This optimal solution is then integrated in time and shown to display a qualitative similarity to the so-called 'minimal defect', for the same parameters. Finally, numerical simulations of a 'edge state' are conducted, to identify an unstable solution which mediates laminar-turbulent transition and relate it to results of the optimisation procedure.
\end{abstract}

\section{Introduction}

Transition to turbulence in ducts is still a challenging issue despite the 125 years since the seminal paper by Osborne \cite{Reynolds_1883} which led to the definition of a dimensionless number, that came some time afterwards to bear his name, capable of broadly discriminating cases of 'streamlined' flow from cases where 'sinuous motion' prevailed.  Osborne Reynolds recognized also that there was no unique value of this dimensionless parameter, representing the ratio of convective to viscous forces, separating the two classes of motion, and that the end state was influenced by the background perturbations present.  The problem of laminar-turbulent transition in shear flows was posed, to fascinate and attract the attention of thousands of researchers in the years to come. Incidentally, the story of how the paper was reviewed by two referees of the caliber of Lord Rayleigh and Sir George Stokes makes for instructive reading \citep{Jackson_ARFM_2007}.

In a subsequent paper, \cite{Reynolds_1895} goes beyond transition and presents, for the first time, the decomposition of the turbulent field into a mean and a fluctuating part, arriving at the equations now known as the 'Reynolds-averaged Navier-Stokes equations', with unknown turbulent stress terms in the mean flow equations. Reynolds had the intuition that a balance is necessary to maintain turbulence, namely production, i.e. the transfer of kinetic energy from the mean flow to the fluctuations, must equilibrate dissipation of the fluctuations, for turbulence to subsist. 

While some classical flows, such as that between differentially heated parallel plates in the gravity field or that between concentric differentially rotating cylinders, exhibit a smooth progression to increasingly complicated patterns via a sequence of bifurcations, whose initial points are well predicted by (modal) linear stability theory, the laminar square duct and the pipe flows are linearly asymptotically stable \citep{Gill_JFM_1965, Tatsumi_JFM_1990}. On the other hand, experiments clearly show that these flows undergo transition abruptly at moderate values of the Reynolds number. The transition to turbulence originates from a subcritical instability which requires perturbations of finite amplitude to bring the flow out of the basin of attraction of the laminar state. Current understanding ascribes the failure of classical stability theory to its focus uniquely on the least stable eigenmode. When a small disturbance composed by a weighted combination of linear eigenfunctions is considered, because of the non-normality of the linearized stability operator, there is the potential for very large transient amplification of the disturbance energy, even in nominally stable flow conditions. This property has been beautifully described in a paper by \cite{Trefethen_Science_1993}, where the authors also show that non-normal operators render the modal instability of pseudo-modes (i.e. modes of a perturbed operator) possible. The definition of pseudospectrum has later been extended to the case of base flow uncertainties by \cite{Bottaro_JFM_2003}.

A recent new direction in rationalising the abrupt transition of many wall-bounded shear flows focuses on identifying alternative solutions (beyond the laminar state) to the governing Navier-Stokes equations. In the last two decades such coherent solutions, generally unstable and in the form of steady states or travelling waves, have been found, first for channel flows \citep{Nagata_JFM_1990, Clever_JFM_1992, Cherabilili_JFM_1997, Waleffe_PoF_1997} and more recently in a pipe \citep{Faisst_PRL_2003, Wedin_JFM_2004}. The solutions computed have the same basic structure, which appears to be in relation to the self-sustaining process (SSP) observed in the near wall region of turbulent shear flows. The three main ingredients of the SSP are: {\it streamwise rolls} of amplitude $1/Re$ which induce  {\it streaks} (an order one spanwise modulation of the ideal laminar flow) by lift-up effect and, to close the loop, a  { \it travelling wave} (an unstable eigenmode of the streak) whose quadratic interactions feed onto the rolls. 

These alternative solutions to the Navier-Stokes equations, also labelled 'exact coherent structures', have been computed by continuation methods. First the rolls are generated by a thermal body force \citep{Clever_JFM_1992}, by a centrifugal force \citep{Nagata_JFM_1990}, or by an unphysical \emph{ad hoc} body force \citep{Waleffe_PoF_1997}. The solution is then obtained by progressively decreasing the forcing imposed to create the rolls. Flow states similar to those computed by continuation approaches have been observed in numerical simulations of turbulent flows. By reducing the size of the computational box,  \cite{Hamilton_JFM_1995} have found the minimal flow unit, containing just a single pair of streaks, capable of maintaining some turbulent activity. Periodic solutions have also been used to describe the bursting event, i.e. the break-up and re-creation of coherent structures such as streaks, in the turbulent buffer layer. \cite{Kawahara_JFM_2001} have found two periodic solutions in plane Couette flow which represent quiescent and turbulent phases of the flow. The phenomenon of bursting has thus been interpreted in phase space as the wandering of the flow trajectory back and forth between such solutions.

Tracking the bifurcation of SSP states by continuation with the Reynolds number has led to two kinds of solutions: the so-called upper and lower branch states.  From a dynamical system point of view, the lower branch solutions sit on a separatrix, the phase-space boundary between the basin of attraction of the laminar flow and that of the turbulence. The lower branch flow is also called an 'edge state'. An open question concerns the number of different states embedded in the separatrix. These exact coherent solutions, with their stable and unstable manifolds, are typically low-dimensional saddle points which collectively produce a chaotic repellor. This leads to the perspective of representing the chaotic dynamics of turbulence through a projection onto a (hopefully not too large) set of 'exact' coherent solutions. The idea is that these structures, connected through their stable and unstable manifolds, are capable to support chaotic trajectories in phase space. A chaotic behaviour with a finite number of degrees of freedom could be explained by the existence of a strange attractor \citep{Ruelle}. The appealing aspect is that turbulence could be described by a finite number of coherent solutions, which constitute the skeleton around which the chaotic dynamics is organised. However, the question of whether a global attractor of the Navier-Stokes equations in three-dimensions exists is an issue that is not yet settled. Beyond its fundamental character, this question has important practical implications, for knowing that a finite dimensional attractor exists would guarantee that the long-time behavior of the Navier-Stokes solutions can be approximated by numerical means  (i.e. using a finite number of degrees of freedom). 

\begin{figure}[!ht]
\begin{center}
\includegraphics[width=12cm]{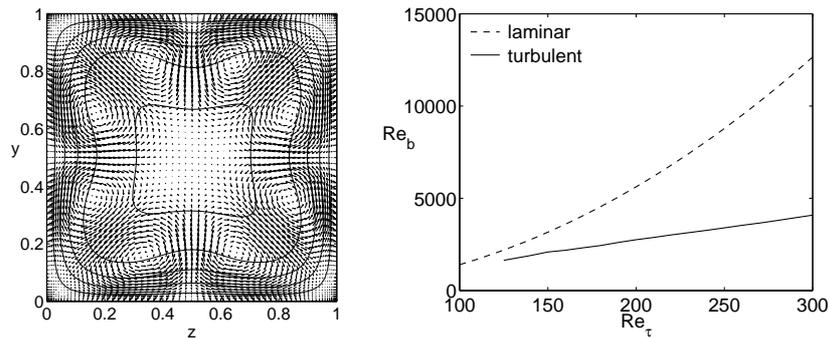}
\caption{\label{turb} Left: turbulent mean crossflow vortices and streamwise flow contours. Spacing between adjacent isolines is equal to $4$ in units of $~ u_\tau$. Right: variation of $Re_b$ with $Re_{\tau}$ in the laminar and turbulent regimes.}
\end{center}
\end{figure}

The present paper deals with the flow through a duct of square cross-section with constant pressure gradient $dP_0/dx$ in the streamwise direction.  While sharing with the cylindrical pipe flow many characteristics (both flows are linearly stable, for example, and undergo spontaneous transition at comparable values of the Reynolds number), the square cross-section, by its geometric features, has the capacity of strongly constraining secondary flows, both instantaneous and time-averaged. An example of this is provided in figure \ref{turb} (left frame) where turbulent mean velocity components arising from a direct numerical simulation at $Re = u_\tau h/\nu = 150$ in a duct of (dimensional) length equal to $6 \pi h$ are displayed, and the presence of symmetric secondary vortices of Prandtl's second kind is shown.  The cartesian coordinates employed in the following are $x$, $y$ and $z$ to define, respectively, the streamwise, vertical and spanwise directions. The scales used to normalise the Navier-Stokes equations are the friction velocity $u_\tau = \sqrt{(-h/4 \rho) dP_0/dx}$, the side length of the square $h$, the density $\rho$ and the kinematic viscosity $\nu$.

Figure \ref{turb} (right), obtained from a series of direct numerical simulations at progressively varying values of the pressure gradient, shows that in the turbulent regime the Reynolds number $Re_b = U_b h /\nu$, $U_b$ bulk velocity, grows almost linearly with $Re$ (noted $Re_\tau$ in this figure for immediate distinction
from $Re_b$), in agreement with results by \cite{Uhlmann_JFM_2007}.  The smallest $Re$ for which turbulence exists is found to be $130$, corresponding to $Re_b\approx 1730$.  This value is slightly smaller than that identified by \cite{Uhlmann_JFM_2007} ($Re=160,~ Re_b=2154$).  \cite{Uhlmann_JFM_2007} have interpreted 
$Re = 160$ as the value defining a 'minimal channel' capable of some form of turbulent activity, with two streaks, each flanked by a pair of quasi-streamwise vortices, active over two walls which face one another.  Such a state with two pairs of rolls adjacent to opposing walls, with the {\it active} role of the walls alternating in time, exists near marginal conditions, and it has been interpreted as the signature of the edge state \citep{Biau_JFM_2008}. Interestingly, the effect of the long time averaging of the two possible sets of marginal states (with the two pairs of rolls lying on the horizontal or the vertical walls) is that of producing a eight-vortex pattern such as that displayed in figure \ref{turb} (left).

The very first step of the bypass transition process in boundary layers, channels and ducts, has often been described as an algebraic instability of the base flow, leading to the formation of streamwise-elongated streaks. Subsequent steps in the process are the linear wavy instability of the streaks and the nonlinear feedback onto the rolls. Despite the fact that streaks, rolls and waves are all present in the exact coherent solutions briefly cited above, there is as yet no description of how such states might arise from the stable laminar base flow.  A path to transition relying on the amplification of so-called optimal disturbances and their subsequent nonlinear development has proven to be sub-optimal, when viewed from the point of view of the minimal initial energy needed to provoke transition \citep{Biau_JFM_2008}.

A different optimization strategy is thus called for, to identify the optimal path from the laminar to the turbulent state, relying on recent developments on the SSP and on physical understanding of the breakdown process.

\section{The optimisation approach}

Evidence collected indicates that key to transition is the establishment of a distorted mean flow profile, capable of supporting strong instabilities. This distortion is different from that caused by the rolls and the streaks issued from classical optimal perturbation theory. 
The recent theory of 'minimal defects' \citep{Bottaro_JFM_2003,Biau_PoF_2004,BenDov_PRL_2007,BenDov_JFM_2007}, yielding the base flow deformation of fixed amplitude capable of maximal destabilisation, is also not sufficient to explain available observations, since it provides no link between the initial state and the deformed flow.

Thus, we have set up a new optimisation strategy, aimed at finding the travelling wave of minimal amplitude capable of yielding a self-sustained state, based on the following initial steps:
\begin{itemize}
\item linear algebraic growth of the travelling wave,
\item generation of weak streamwise vortices by quadratic interactions,
\item production of a strong streak by lift-up,
\item regeneration of the travelling wave, closing the loop for a sustained state.
\end{itemize}

Velocity and pressure are decomposed into a base, laminar state $(U_0, P_0)$, and a time-dependent part. The time-dependent component is itself decomposed into a slowly time-varying mean flow defect ($Q$) plus a streamwise travelling wave ($q$) of order $\mathcal{O}(\epsilon) \ll 1$. The vortex $(V,~ W)$ is of 
order $\mathcal{O}(\epsilon^2)$, while the streak $U$ resulting from lift-up is $\mathcal{O} (\epsilon^2 Re)$.  Using the decomposition:
$$ \left[ \begin{array}{l} U_0(y,z) \\ 0 \\ 0 \\ P_0(x) \end{array} \right] + 
\left[ \begin{array}{l} U(y,z,t) \\ V(y,z,t) \\ W(y,z,t) \\ P(y,z,t)  \end{array} \right] + 
\left[ \begin{array}{l} u(x,y,z,t) \\ v(x,y,z,t) \\ w(x,y,z,t) \\ p(x,y,z,t) \end{array} \right], $$
the equations governing the mean, streamwise-averaged flow read:
\begin{eqnarray*}\label{RANS}
V_y + W_z &=&0, \\
U_t + V(U_0+U)_y + W(U_0+U)_z &=&        \frac{1}{Re}( U_{yy}+U_{zz} ) - \left(\overline{vu}|_y+\overline{wu}|_z \right),\\
V_t                           &=& -P_y + \frac{1}{Re}( V_{yy}+V_{zz} ) - \left(\overline{vv}|_y+\overline{wv}|_z \right),\\
W_t                           &=& -P_z + \frac{1}{Re}( W_{yy}+W_{zz} ) - \left(\overline{vw}|_y+\overline{ww}|_z \right),
\end{eqnarray*}
associated to homogeneous boundary conditions $U=V=W=0$ on the walls. Note that in the equations above overbars over the products of fluctuating velocity components denote averaging along the streamwise distance. The Navier-Stokes equations, linearized around the streaky flow ($U_0+U$), are:
\begin{eqnarray*}\label{LNS}
u_x +v_y+w_z &=&0, \\
u_t +(U_0+U) u_x + v(U_0+U)_y+w(U_0+U)_z &=& -p_x + \frac{1}{Re}( u_{xx} + u_{yy} + u_{zz} ),\\
v_t +(U_0+U) v_x                         &=& -p_y + \frac{1}{Re}( v_{xx} + v_{yy} + v_{zz} ),\\
w_t +(U_0+U) w_x                         &=& -p_z + \frac{1}{Re}( w_{xx} + w_{yy} + w_{zz} ),
\end{eqnarray*}
together with $u=v=w=0$ on the walls and streamwise periodicity.

The energies of mean flow defect ($E$) and fluctuations ($e$), to be used later, are defined as:
$$E(t) = \frac{1}{2}\int_{yz} (U^2+V^2+W^2) ~ dy~dz,$$
and
$$e(t) = \frac{1}{2}\int_{xyz} (u^2+v^2+w^2) ~ dx~dy~dz.$$

A gain is defined as $G = e(t)/e(0)$, and the goal of the work is to find the initial wave at $t=0$ capable of maximising $G$.  In the first step of the iteration procedure, $U, V$ and $W$ are zero; they are created from the second iteration on, because of the presence of the Reynolds stress terms in the mean flow equations. As an alternative it might have been interesting to test the approach with a different objective functional, such as a gain based more directly on the Reynolds stresses or the ratio between production and dissipation terms.

The optimisation procedure is standard and, instead of carrying out a constrained optimisation, we introduce a Lagrangian functional $\mathcal{L}$ to be maximised without constraints. This augmented functional is:
$$ \mathcal{L} = G - <Q^\dag,~ L(Q,~q)> - <q^\dag,~ l(Q,~q)>, $$
where $<.>$ denotes the usual inner product (i.e. volume integration of the product of the two quantities over the domain) and $ Q^\dag , q^\dag $ are Lagrange multipliers. On an extremum point, the following necessary conditions must be satisfied:
$$ \frac{\partial \mathcal{L}}{\partial Q} \delta Q = 0 \quad \mathrm{and}\quad 
   \frac{\partial \mathcal{L}}{\partial q} \delta q = 0,$$
leading to the adjoint problems in symbolic form:
$$ < L^\dag Q^\dag,~ \delta Q> = - <q^\dag,~ l(\delta Q)q> = <G_{Q},~ \delta Q>, $$ and
$$ < l^\dag q^\dag,~ \delta q> = - <Q^\dag,~ L(\delta q)Q> = <g_{q},~ \delta q>, $$
(superscript $\dag$ designs adjoint quantities) which must be solved together with the direct equations for the fluctuations and the mean flow; $\delta Q$  and $\delta q$ are infinitesimal variations of the mean flow and of the fluctuations, $ G_{Q}$ and $ g_{q}$ are sensitivity functions which arise naturally after integration by parts when isolating $\delta Q$  and $\delta q$.

In extended form, the adjoint equations for the mean flow are:
\begin{eqnarray*}\label{LNSadjointM}
V^\dag_y+W^\dag_z&=&0, \\
-U^\dag_t                    &=&             \frac{1}{Re}( U^\dag_{yy}+U^\dag_{zz} ) ~ + ~ G_{U},\\
-V^\dag_t + U^\dag (U_0+U)_y &=& -P^\dag_y + \frac{1}{Re}( V^\dag_{yy}+V^\dag_{zz} ),\\
-W^\dag_t + U^\dag (U_0+U)_z &=& -P^\dag_z + \frac{1}{Re}( W^\dag_{yy}+W^\dag_{zz} ),
\end{eqnarray*}
with $U^\dag=V^\dag=W^\dag=0$  on the walls. The adjoint equations for the fluctuations are:
\begin{eqnarray*}\label{LNSadjointF}
 u^\dag_x +v^\dag_y+w^\dag_z&=&0, \\
-u^\dag_t-(U_0+U)u^\dag_x \quad           &=& -p^\dag_x + \frac{1}{Re}( u^\dag_{xx}+u^\dag_{yy}+u^\dag_{zz} )~+~ g_{u},\\
-v^\dag_t-(U_0+U)v^\dag_x+u^\dag(U_0+U)_y &=& -p^\dag_y + \frac{1}{Re}( v^\dag_{xx}+v^\dag_{yy}+v^\dag_{zz} )~+~ g_{v},\\
-w^\dag_t-(U_0+U)w^\dag_x+u^\dag(U_0+U)_z &=& -p^\dag_z + \frac{1}{Re}( w^\dag_{xx}+w^\dag_{yy}+w^\dag_{zz} )~+~ g_{w},
\end{eqnarray*}
with $u^\dag=v^\dag=w^\dag=0$ on the walls and periodic boundary conditions along $x$. Note the minus signs in front of the time derivatives of the 'adjoint momentum equations', indicating that the only possible direction of stable evolution is negative time. All the adjoint equations are linear but are coupled to the direct state, and this represents a numerical challenge since the direct fields must be stored at all time steps.

The sensitivity terms are:
\begin{eqnarray*}\label{Sensitivity}
 G_U &=& -u^\dag (u_x+v_x+w_x)_x + (u^\dag v)_y + (u^\dag w)_z,  \\
 g_{u} &=& -U^\dag_y v-   U^\dag_z w,\\
 g_{v} &=& -U^\dag_y u- 2 V^\dag_y v + w(V^\dag_z+W^\dag_y),\\
 g_{w} &=& -U^\dag_z u- 2 W^\dag_z w + v(V^\dag_z+W^\dag_y).
\end{eqnarray*}
\noindent $G_U$ corresponds to the sensitivity to variations in the base flow; it is a generalisation of the function found by \cite{Bottaro_JFM_2003}. The terms $g_u,~ g_v$ and $g_w$ correspond to wave sensitivity functions, and are necessary for the optimisation of the feedback loop. 

A sequence of direct and adjoint calculations is carried out to optimise $G$
by employing spectral collocation codes which are slight modifications
of those employed by \cite{Biau_JFM_2008}.
 The result of the integration of the direct system up to $t = T$ (where $T$ is the target time of the optimisation) provides $q(T)$ (and $Q(T)$); the adjoint system is integrated backward in time and is initialised by  $q^\dag (T) =  q (T)$ and $Q^\dag (T) = 0$.  Once the adjoint fields are computed, the direct calculations are re-initialised with $q (0) =  q^\dag(0)$ and $Q(0) = 0$, so that the cycle can restart.  At convergence we obtain the optimal initial fluctuating field
$q(0)$ whose energy is maximised at $t=T$, and the associated base flow distortion $Q(t)$ produced by the action of the fluctuations onto the mean field. 

The disturbances are expressed using a single-mode Fourier decomposition in the streamwise direction, i.e.:
$$q(x,y,z,t) = \tilde{q}(y,z,t)e^{+i\alpha x}+\tilde{q}^*(y,z,t)e^{-i\alpha x},$$
where superscript $*$ denotes complex conjugate. The present representation is acceptable if the first stages of the transition process are focussed upon.

\section{Optimisation results}
\begin{figure}[!ht]
\begin{center}
\includegraphics[width=7cm]{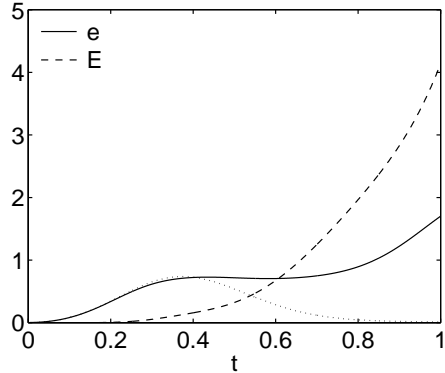}
\caption{\label{NRJ} Behaviour in time of the energy $e$ of the optimal initial fluctuations and of the energy $E$ of the optimal defect. The dotted line is the result of a simulation of the linear disturbance equations in the absence of mean flow distortion, starting from the optimal initial disturbance field.}
\end{center}
\end{figure}

In the following, the Reynolds number is fixed at the value $Re=150$. Figure \ref{NRJ} shows results of the optimisation procedure for the case $\alpha=1$, $e_0=3\times10^{-3}$, and for a final target time $T=1$.  Initially, the transient behaviour of the fluctuation energy follows the linear curve (shown with a dotted line) obtained in the absence of a base flow defect.  Within the optimal procedure developed here the energy $E$ of the mean flow distortion increases, as a result of Reynolds stress feedback. For $t$ larger than about $0.8$  the deformed mean flow can sustain a rapid amplification of the fluctuations, which brings the disturbance energy $e$ to a level unattainable by the transient growth process alone. Thus it appears that transient amplification and defects concur in defining the initial stages of transition, as postulated by \cite{Biau_PoF_2004}.  

\begin{figure}[!ht]
\begin{center}
\includegraphics[width=12cm]{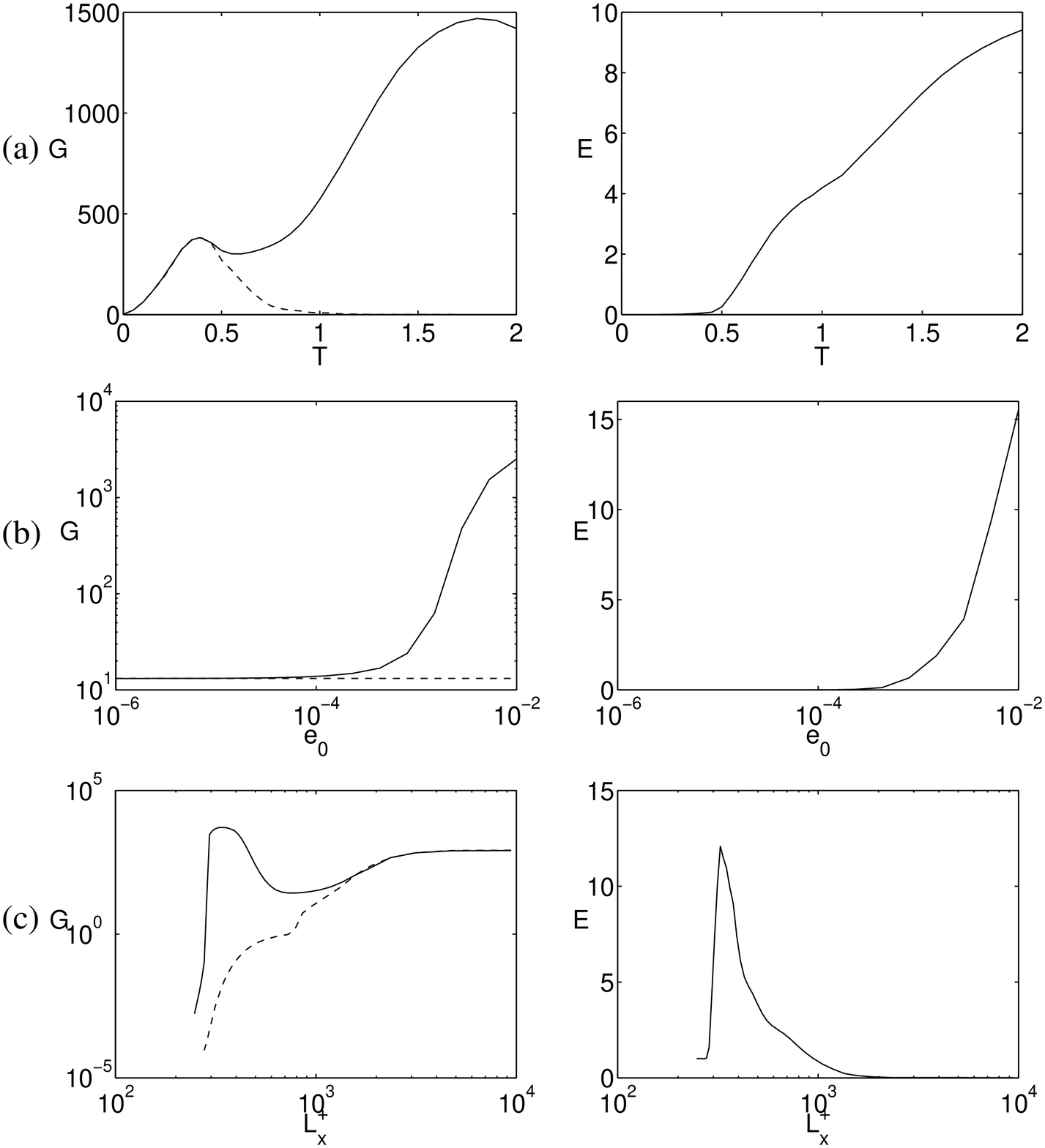}
\caption{\label{param} Parametric study. (a) $\alpha=1;~ e_0=3\times 10^{-3}$. (b) $\alpha=1;~ T=1$. (c) $T=1;~ e_0=10^{-3}$. The dashed lines represent the linear case, in the absence of mean flow defect.}
\end{center}
\end{figure}

Figure \ref{param} reports results from a detailed parametric study of the problem, under the conditions indicated in the caption of the figure. It should be preliminarily be observed that gains of order $800$ can be achieved by the transient amplification mechanism in the linear case when $\alpha = 0$ \citep{Biau_JFM_2008}.  However, for $\alpha = 0$ no mean flow distortion is produced in a nonlinear setting, and at large times the disturbances simply decay.  For $\alpha = 1$ the results are more interesting.

The two frames labelled as (a) show that for $T$ sufficiently long both $e$ and $E$ can increase very much.  Within the time interval considered, the energy gain of the fluctuations is optimal for $T \approx 1.8$, i.e. over a physical viscous time scale of order $h/u_{\tau}$, which is the characteristic time of evolution of the streaks.  At the imposed value of $e_0$, the defect plays a role only for $T$ larger than about $0.5$; the result of optimisations runs carried out for smaller target times coincide with classical optimal perturbation results (dashed curve). 

The effect of varying $e_0$ is displayed in frames (b); as a result of nonlinearities the gain can increase very much, causing subcritical transition when the initial disturbance energy level is sufficient.  

The effect of the computational box size is represented in frames (c).  There are two noteworthy results here.  The first is that there appears to be a clear cut-off length $L^\dag_x \approx 250$, below which the cycle is not self-sustained. This is the 'minimal box dimension' for transition to appear, and the result is consistent with direct simulations by \cite{Uhlmann_JFM_2007}.  The second interesting point is that, for $T = 1$, there is an optimal box length, equal to $L^\dag_x \approx 350$ (the corresponding wavenumber is $\alpha = 2 \pi Re / L^\dag_x \approx 2.7$), where maximum amplification of both fluctuations and mean flow defect is achieved.  These results are mitigated by the observation that a single wavenumber disturbance has been considered in the model, as already pointed out earlier.  In reality, one should expect streamwise-Fourier modes interactions when long computational boxes are numerically computed.  When a single mode with $\alpha$ very small is considered (i.e.  $L^\dag_x$ larger than 2500) the results obtained here are superposed to the (classical) optimal case; the amplification process coincides with the phenomenon of lift-up of low speed streaks from the wall, with a base flow which is not distorted from the ideal conditions.  This demonstrates that, although classical optimal perturbation theory \citep{Butler_PoF_1992} yields initial flow states capable of large growth for small $\alpha$, these are inefficient at triggering the sequence of processes which eventually causes breakdown of the flow. The reason is that no defect can be generated by the Reynolds stress terms, as shown in frame (c) (right) of figure \ref{param}. The optimisation strategy outlined in this paper, while recovering results of the classical theory for $\alpha$ small, highlights the importance of closing the feedback loop.

\section{The minimal defect}
The previous section demonstrates the importance of mean flow defects for the sustainement of fluctuations. It is thus interesting here to introduce a link between optimal perturbations and minimal defects, \emph{i.e.} those base flow distortions of minimal energy capable to induce efficiently subcritical instability \citep{Bottaro_JFM_2003}.
For transitional flows (i.e. low amplitude mean flow distortions), the concept of minimal defect provides a new outlook on the possibility of rapid disturbance growth. For turbulent flows (distortions of finite amplitude), these defects display characteristics of possible relevance to the dynamics of coherent eddies.

The following is a brief summary of the methodology to compute minimal defects. Let us consider the linear stability equations in symbolic form as an eigenvalue problem: 
$$L(U_0)q = -i \omega q.$$ 
A steady base flow defect $\delta U$ induces variations on the perturbations and on the corresponding eigenvalues, i.e.
$$U_0+\delta U \Rightarrow 
\left\lbrace \begin{array}{l}
\omega + \delta \omega \\
q+\delta q
\end{array}\right.$$
The linear stability problem then becomes:
$$L(U_0+\delta U) (q+\delta q) = -i (\omega+\delta \omega) (q+\delta q),$$
the first variation of which is
$$ L(U_0+\delta U) q + L(U_0) \delta q = -i \omega\delta q-i\delta\omega q.$$

\begin{figure}[!ht]
\begin{center}
\includegraphics[width=13cm]{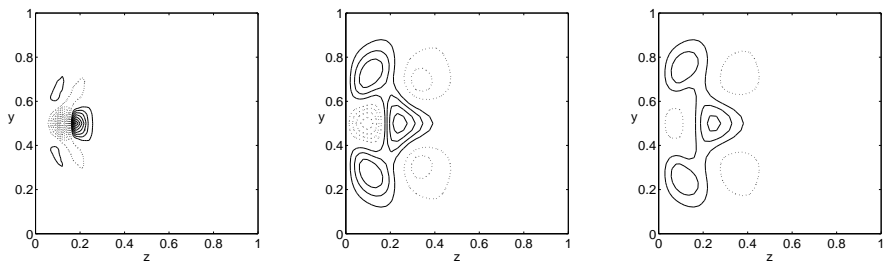}
\caption{\label{MinDef} Left frame: the minimal defect is plotted with isolines of the streamwise disturbance velocity, for $E_\Delta =0.6598$ and $\alpha = 1$. Center frame: snapshot of the streak $U$ obtained by the optimisation procedure of section 2.  The figure corresponds to $t=0.6$ in figure \ref{NRJ}. Right frame: instantaneous result ($t=0.6$) of a direct simulation of the transition process (corresponding to the continuous black line of figures \ref{skin_friction} and \ref{PhaseDiag}) initialised with the optimal solution of figure \ref{NRJ}. In all three frames the spacing between the $U$ isolines is equal to one (i.e. $u_\tau$ in dimensional terms), and the dotted lines denote negative values.}
\end{center}
\end{figure}

\noindent In order to isolate the variation of $\omega$, we use the adjoint variable $q^\dag$ defined by:
$$<q^\dag,~ (L+i\omega)q>=<q,~ (L^\dag-i\omega^*) q^\dag>=0,$$
with an appropriate inner product \citep{Bottaro_JFM_2003}. We can thus obtain the sensitivity function $G_U$:
$$\begin{array}{rcl}
\delta \omega & = i & <q^\dag,~ L(U_0 + \delta U) q > \\
              & = & <G_U,~ \delta U>
\end{array}$$
with the normalization $ <q^\dag,~ q>=1$. Note that here the function $G_U$ differs from that obtained earlier. The minimal defect is found by maximising the growth rate $\omega_i = Imag(\omega)$, under the constraint that the energy of the defect $\int_{yz} (U-U_0)^2 ~ dy~ dz$ is fixed at the value $E_\Delta$. This latter constraint is a simplified state equation for the base flow deviation. The Lagrangian functional takes the form:
$$ \mathcal{L} = \omega_i - \lambda \left(\int_{yz} (U-U_0)^2 ~ dy dz - E_\Delta \right),$$
and, by imposing that $({\partial \mathcal{L}}/{\partial U})~\delta U=0$, the optimisation loop is:
$$\left\lbrace 
\begin{array}{l}
\lambda^{n+1} = \sqrt{ \int_{yz}  Imag(G_U^{n})^2~ dydz /4 E_\Delta }\\
\delta U^{n+1} = Imag(G_U^n) / 2 \lambda
\end{array}
\right.
$$
\noindent with the exponent $n$ denoting the iteration number. An illustrative result is presented in figure \ref{MinDef} (left frame). The qualitative similarity between the results in the left frame and that on the central frame (corresponding to the optimisation of section 2) is clear, with a low speed streak at the center of wall flanked by two high speed streaks, although the minimal defect displays much stronger gradients of the velocity and presents a much larger (local) instability growth rate.  It has to be stressed the fact that the minimal defect is a local feature which does not satisfy the momentum conservation equation. When it is imposed as initial condition in a direct simulation, the sharp gradients diffuse rapidly leading to a solution which resembles more to that displayed in the two right frames. The figure in the right frame of figure \ref{MinDef} is very similar to that arising from the optimisation procedure, aside from the smaller peak absolute values of $U$ attained. In the direct numerical simulation on the right frame of figure \ref{MinDef} nonlinearities contribute to a redistribution of the energy among many flow harmonics, whereas in the central frame the only wavenumbers which can accommodate the available energy are $\alpha=0$ and $\alpha=1$. Despite this, the agreement is sufficiently good to provide confidence in the ability of our simple model to provide correct flow features in the initial phases of transition.  It is also interesting to observe that the flow patterns displayed in figure \ref{MinDef} resemble the asymmetric solution computed by \cite{Pringle_PRL_2007} for the case of the pipe.   

An open question concerns the mechanism which associates the position of each streamwise vortex to its sign of rotation. \cite{Uhlmann_JFM_2007} propose a heuristic explanation based on a simple kinematic analysis of the interaction of streamwise vorticity with a corner. The interaction of a streamwise vortex, close to one of the corners, with the impermeable wall can be modelled using three image vortices and potential theory. Thus the vortex would migrate because of the induced velocity field towards a stable position. This argument is consistent with observations. Considering the results in figure \ref{MinDef} another hypothesis can be formulated: a low speed streak sitting on a bisector is the 'most active' flow feature, i.e. it is that particular flow structure capable of better sustaining the wall cycle.

If we express the wavy perturbations as $q(x,y,z,t)=A(t)\hat{q}(y,z)exp(i\alpha x-i\omega_r t)$, the variation in time of the amplitude, when a mean flow is mildly distorted, obeys the equation:
$$\frac{1}{A}\frac{dA}{dt} = \omega_i+\delta \omega_i = \omega_i + <Imag(G_U),~ \delta U>.$$
The product $<Imag(G_U),~ \delta U>$ can be interpreted as a Landau coefficient which determines the nonlinear character of the bifurcation; if this coefficient is positive then the nonlinearity is destabilising and the bifurcation is subcritical. It is interesting here to note that $\delta U$ is always proportional to $A^2 Re$ whether we consider the lift-up regenerating loop (through the formation of streamwise vortices) or the direct generation of streaks by the Reynolds stresses in the streamwise momentum equation (and the latter assertion is clear after realising that the characteristic time scale of evolution of the defect is the viscous scale).  It is thus feasible to assume that a second self-sustaining mechanism exists, with the streaks directly re-generated by the fluctuations.

\section{Nonlinear simulations of the edge state}

\begin{figure}[!ht]
\begin{center}
\includegraphics[width=13cm]{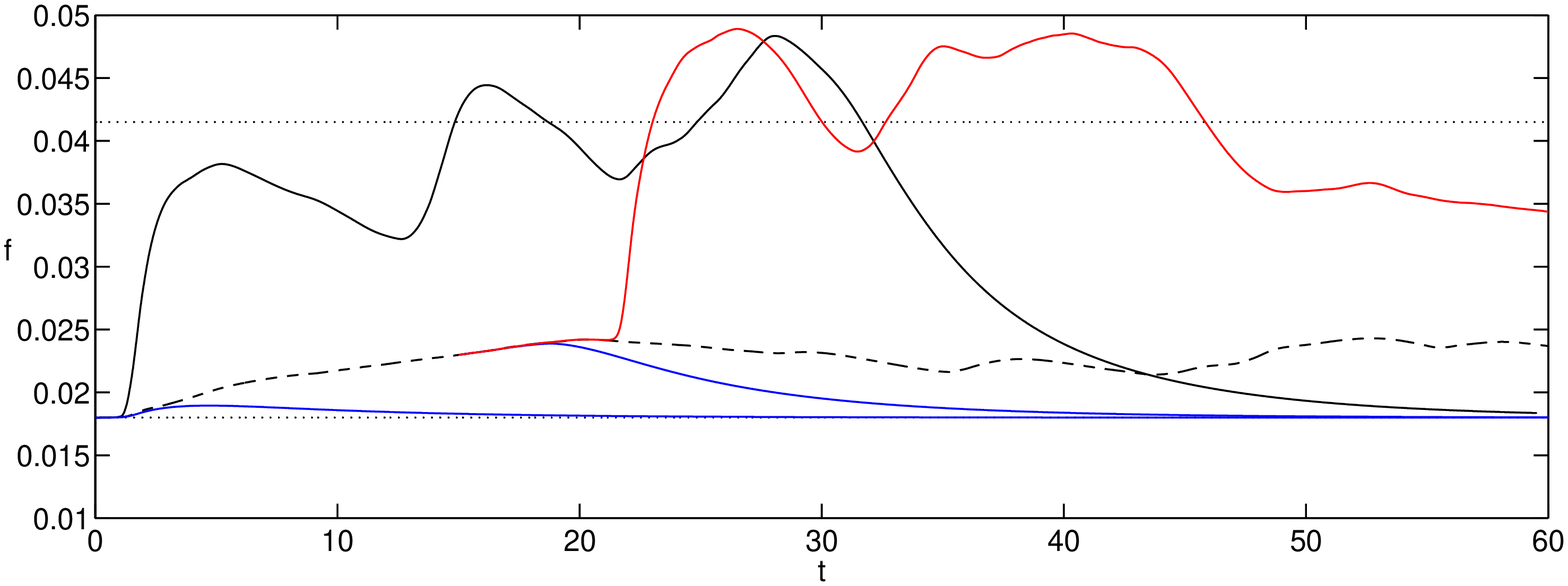}
\caption{\label{skin_friction} Skin friction versus time. Dotted lines indicate the laminar and the (mean) turbulent values of the skin friction. The dashed curve corresponds to the edge state. Blue trajectories relaminarise, the black and the red trajectories lead to turbulence (before eventual relaminarisation at large times).}
\end{center}
\end{figure}

Numerical simulations of non-trivial states are described in this section; the calculations at $Re = 150$ are initialised at $t=0$ with the laminar flow plus the optimal wavy perturbation of figure 2 (with $\alpha=1$
and $e_0=3\times 10^{-3}$) multiplied by an appropriate scaling factor $\beta$. The time evolution of the skin friction is depicted in figure \ref{skin_friction} and the blue trajectory which relaminarises after a few units of time has $\beta = 0.92$. The edge of chaos is a stable manifold residing in phase space somewhere between the laminar and the turbulent flows.  To find this invariant object the procedure described by \cite{Schneider_PRL_2007} has been adopted, successively refining (every viscous time unit) the initial guesses on either sides of the edge surface in phase space, through a continuous update of the relaxation parameter $\beta$.  Four intermediate solutions (a couple initiated at time $t=0$ and a couple initiated at $t=15$, on both sides of the edge) are also shown with colours in figure \ref{skin_friction}, highlighting the exponential separation with $t$ within each pair, with the trace of one state swinging up to turbulence and the other relaminarising.

\begin{figure}[!ht]
\begin{center}
\includegraphics[width=13cm]{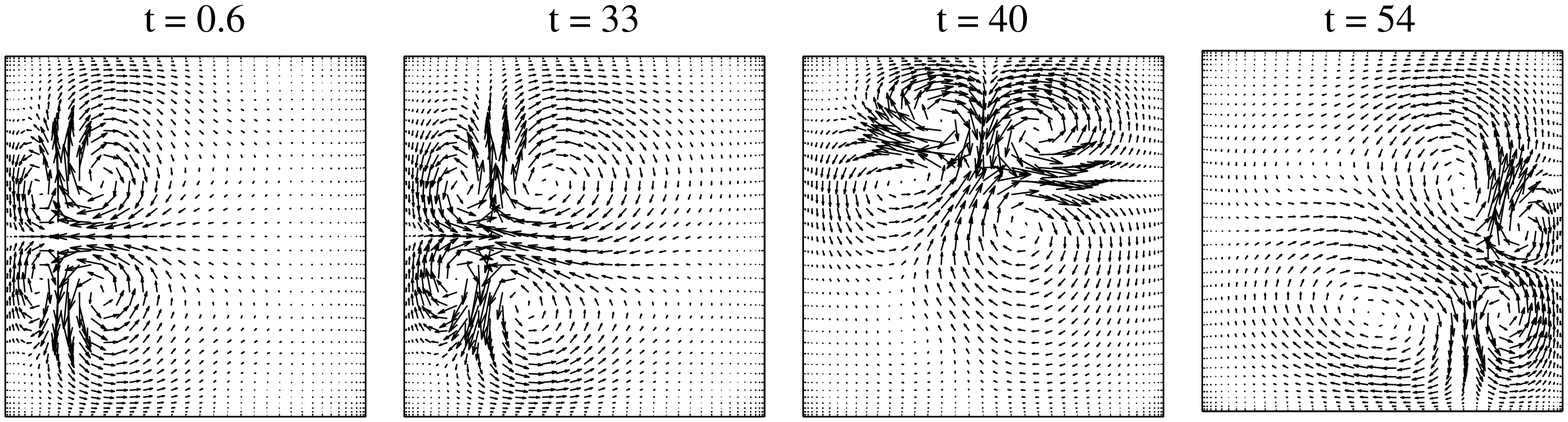}
\caption{\label{EdgeState_T} Instantaneous secondary flows of the solution on the edge.}
\end{center}
\end{figure}

The edge state lives in between and is represented by the dashed line.  Snapshots of the secondary flow patterns along the dashed curve are provided in figure \ref{EdgeState_T}. It is interesting to observe that the basic building block of the edge state is represented by two pairs of vortices. The larger outer pair is weaker, it sits above a smaller near-wall pair, which presents an upwash region fluctuating around a bisector.  The patterns displayed in the figure
are similar to those shown in figure 4; here, as time varies, different walls become 'active'.  The time-averaged edge state in a pipe of circular cross-section consists of one strong pair of vortices near the surface with a much weaker pair above it \citep{Schneider_PRL_2007, Eckhardt_ARFM_2007, Pringle_PRL_2007}. 

\begin{figure}[!ht]
\begin{center}
\includegraphics[width=13cm]{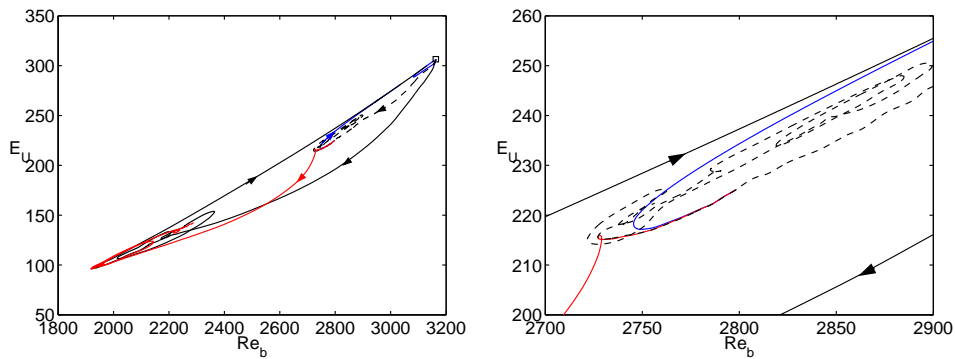}
\caption{\label{PhaseDiag} $E_U$ 'energy' versus $Re_b$. The laminar fixed point with $Re_b=3163$ and 
$E_U=306.45$ is denoted by the square symbol.  The figure on the right frame provides better details
of the flow trajectory on the edge (dashed curve).}
\end{center}
\end{figure}

In turbulent shear flows, the spatio-temporal chaos is extensive, i.e. the number of degrees of freedom scales with the system's volume. Thus it is impossible to represent a turbulent flow as a dynamical system in a low dimensional phase diagram. \cite{Toh_JFM_2003} used the total energy input versus the dissipation rate as a two-dimensional picture of the dynamics. Here, in figure \ref{PhaseDiag}, we use a reduced phase diagram representation spanned by the energy of the streamwise-averaged flow, $E_U=\frac{1}{2}\int_{yz}(U+U_0)^2+V^2+W^2 ~ dydz$, and by the bulk Reynolds number (see also \cite{Biau_JFM_2008}).

The black curve, i.e. the trajectory that goes to turbulence shortly after $t=0$, shows that chaos is attained very rapidly (on a convective time scale) once the trace starts diverging from the edge. After a few orbits around an unstable (turbulent) node the flow is ejected along the unstable-laminar manifold. During the relaminarisation process, the patterns of motion remain self-similar while slowly decaying. The $x-$structure disappears first and the flow tends to be formed by quasi-straight rolls and streaks, which then decay on a viscous time scale towards the laminar fixed point. The dashed curve in figure \ref{PhaseDiag} shows the trajectory which oscillates on the edge surface: this state is a relative attractor and all trajectories which are initially on the edge surface remain confined to it.  Conversely, if the initial condition is close to the edge surface, but not exactly onto it (the case of the red and blue curves), 
the trajectory will arrive close to the saddle, before being ejected away. 

It is significant that the optimal initial condition obtained from the procedure outlined in section 2 yields a solution which resembles that found on the edge and follows closely, for some time, the dashed curve in figure \ref{skin_friction}, before departing away from it.  This fact signals that the simplified optimisation procedure that we have devised produces an appreciably good estimate of the 'true optimal' (as also attested by the fact that the relaxation parameter $\beta$ needed to maintain the solution on the edge is close to one), where by 'true optimal' we define that fluctuating state of minimal initial energy $e_0$ capable to provoke transition.  The initial disturbance of minimal energy causing breakdown arrives arbitrarily close to the hyperbolic point, and, of course, the closer it gets, the more slowly transition is triggered.

\section{Conclusions}

Transition in many wall-bounded shear flows arises from a complicated interaction of rolls, streaks and waves; 125 years ago Osborne Reynolds was able to catch a glimpse of this interaction. Efforts since, to unravel the secrets of transition to turbulence, have focused mostly on linear stability theory (modal and nonmodal) and, more recently, on 'exact coherent states', the self-sustaining process and, to a lesser extent, on 'minimal defects'.  We have tried to put together these individual bricks to provide a coherent picture of the process. 

The new optimisation approach described here provides for the first time a complete model of a
transitional path including, in particular, the feedback of the fluctuations onto the mean flow. We have shown that the initial stages of transition rely on a combination of algebraic and exponential amplification of disturbances, with the latter closely associated to the creation of a mean flow defect. If disturbances of long streamwise wavelength are initially excited by the receptivity conditions, mean flow distortions cannot be created, except in the obvious case of very noisy environment.

Albeit simplified, our optimal model of transition has produced a solution which sits initially on a trajectory directed towards the hyperbolic point on the edge surface, i.e. the separatrix between the laminar basin of attraction and the chaotic dynamics. Direct simulations have confirmed the suitability of the model proposed. The basic flow structure of the edge state in the cross-section of the square duct is formed by two overlapping pairs of vortices, with the near-wall pair more intense. This pattern resembles that obtained by minimal defect theory and is also similar to the asymmetric edge state computed by
\cite{Schneider_PRL_2007, Eckhardt_ARFM_2007}, \cite{Pringle_PRL_2007} and \cite{Duguet_JFM_2008} for the pipe flow.

Work in progress focuses on the effects of increasing the Reynolds number, to try and
compute the unsteady coherent solutions encountered during turbulent intermittency or in puffs. 
From a practical point of view, it should be possible to use the optimisation technique introduced here
also to determine semi-empirical transition criteria based on a simple balance between production and dissipation, as indicated by \cite{Reynolds_1895}.

\begin{acknowledgements}
The support provided by the EU under Marie Curie grant EST FLUBIO 20228-2006 is gratefully acknowledged.
\end{acknowledgements}

\bibliographystyle{agsm}
\bibliography{PhilTrans}

\label{lastpage}

\end{document}